\documentclass[aps,showpacs,preprint]{revtex4}
\usepackage{graphicx}

\begin{document}

\title{Interference mechanism of magnetoresistance in variable range hopping conduction: the effect of paramagnetic electron spins
and continuous spectrum of scatterer energies.}

\author{A.\,V. Shumilin}   \email{AVShumilin@mail.ioffe.ru}
\author{V.\,I. Kozub}

\affiliation{Ioffe Physical-Technical Institute of the Russian
Academy of Sciences, 194021, Saint Petersburg, Russia.}

\begin{abstract}
Despite the fact that the problem of interference mechanism of
magnetoresistance in semiconductors with hopping conductivity was
widely discussed, most of existing studies were focused on the
model of spinless electrons. This model can be justified only when
all electron spins are frozen. However there is always an
admixture of free spins in the semiconductor.

This study presents the theory of interference contribution to
magnetoresistance that explicitly includes effects of both frozen
and free electron spins. We consider the cases of small and large
number of scatterers in the hopping event. For the case of large
number of scatterers the approach is used that takes into account
the dispersion of the scatterer energies. We compare our results
with existing experimental data.
\end{abstract}


\maketitle

\section{Introduction}

At low temperatures the conductivity in semiconductors
 is supported by carriers hopping between
localized states on the impurities. At  low enough temperatures
the characteristic act of hopping occurs not between the
neighboring impurities but between the closest impurities in the
thin energy strip around the Fermi level. This phenomenon is known
as variable range hopping conductivity and is characterized by the
well-known temperature dependence of resistance:
\begin{equation} \label{Mott}
R(T) \propto \exp \left(\frac{T}{T_0} \right)^\alpha,
\end{equation}
where $T_0$ is a constant and $\alpha$ is less than unity. When
constant density of states $g(E)$ is discussed, the exponent $\alpha$ is equal to $1/(d+1)$,
  where $d$ is the system dimension. However when the
conduction strip becomes narrower and the Coulomb gap becomes
important, then $\alpha=1/2$  for any system dimension. The variable
range hopping conductivity is, in particular, discussed in details
in \cite{Book}.

The problem of magnetoresistance in the semiconductors has been actual since 1950-th \cite{sladek}.
 Then it was
understood that at strong magnetic fields the magnetoresistance is
positive and can be exponentially strong. This magnetoresistance
is due to electron wavefunction shrinkage in the magnetic field.

However it has appeared that at weaker fields the
magnetoresistance becomes negative. This phenomenon has been first
described by Nguen, Shklovskii and Spivak \cite{NSS} (for detailed
review see \cite{SS}). It is related to the interference. During
the hop the tunneling carrier suffers under-barrier scattering  on
the impurities that are outside of the conduction strip. So the
resulting hopping amplitude becomes the sum of different tunneling
paths, which interfere with each other.

It is important that the hopping conductivity is controlled by
percolation and, thus, by the largest resistors (formed by the
pairs of hopping sites) that still should be included into the
percolation network. This fact strongly emphasizes a role of
destructive interference which can practically cut the percolation
path and leads to an increase of the resistance. The magnetic
field suppresses the interference and leads to  negative
magnetoresistance.

The negative magnetoresistance appears to be linear (on the
magnetic field) in the relatively wide range of weak fields, so it
dominates over quadratic positive wave-shrinkage magnetoresistance
at weak magnetic field.

The theory of interference magnetoresistance \cite{SS} is based
on the model of spinless electrons. This model can be justified
when all electron spins are frozen, for example, by the exchange
interaction. In \cite{SS} it is stated that there is no
interference effect on the magnetoresistance when all electron
spins are free. However this statement lacks the solid theoretical
proof. Also, no discussion of the systems in which some spins are
frozen and some
 are free is given. Therefore the question remains "how many free spins one
should have to suppress the interference magnetoresistance?".

In \cite{Spivak} it is stated that the admixture of free
spins leads to additional positive magnetoresistance due to spin
alignment in the magnetic field. However  no detailed
discussion of this magnetoresistance mechanism is given. For
example its detailed temperature dependence remained unknown.

In the studies \cite{NSS, SS} authors consider the system with a
large number of scatterers involved in a hopping event. In our
study we denote this situation  as the case of long hops. For this
model only qualitative results are obtained in \cite{NSS, SS}.
Later it was shown  \cite{Shirmacher, Raikh} that in many
realistic systems the characteristic number of scatterers is
small. Most hopping events occur without scattering, and to
consider interference magnetoresistance one can take into account
only hopping events including a single scatterer. We denote this
situation as the case of short hops. For this problem of short
hops (and spinless electrons) the papers \cite{Shirmacher, Raikh}
give quantitative estimates of the magnetoresistance.

There were several attempts \cite{PRL_perc,
Medina,PRL_an,an,Spin-Orb} to develop a quantitative theory for
long hops. However, to our opinion, no one of this works pays a
sufficient attention to the variance of scatter energies.

Also we state that the conventional theory of interference contribution to
magnetoresistance is not sufficient to describe existing
experimental data. While semiconductors with hopping conductivity
usually exhibit a combination of negative linear and positive
quadratic magnetoresistance (this fact is in accordance with
conventional magnetoresistance theory), the temperature dependence
of negative and positive magnetoresistance differs sometimes from
predicted by existing theories \cite{Rentzsch, ours1, ours2,
ours3, ours4, ours5, ourPRB}. In \cite{Rentzsch, ours2, ourPRB}
this difference is attributed to spin effects. In \cite{ours2,
ourPRB} the spin alignment magnetoresistance (suggested in
\cite{Spivak}) is invoked to describe experimental observations.
However no rigorous theory of this magnetoresistance (and any
consistent discussion of free spins effect on the
magnetoresistance)is presented.

The goal of the present study is to develop theoretical approach
to the interference contribution to magnetoresistance that
explicitly takes into account an admixture of free spins. We
consider two general problems. The first is to what extent the
interference magnetoresistance is suppressed when spins become
free. The second is related to derivation of the expression for
spin alignment magnetoresistance that would give the detailed
dependence of this magnetoresistance on the system parameters
including magnetic field, temperature and the weight of the free
spins.

In our work we consider both long and short hops. For  long hops
(large scatterer numbers) we use an approach that takes into
account the scatterer energy variance. Our approach  becomes
rigorous for exponentially broad distribution of scatterers
energies. However we show with numerical computations that it is
applicable to realistic energy variances.

The plan of this study is as follows. In section \ref{sect_gen} we
discuss the basics of interference magnetoresistance mechanism. In
section \ref{sect_shorthops} we present theoretical analysis for the
short hop case. In section \ref{sect_longhops1} we discuss the
statistics of tunneling path amplitudes in the long hop limit. In
section \ref{sect_lonhhops2} we use this statistics to get
expression for magnetoresistance in this case. In section
\ref{sect_exp} we compare our results with existing experimental
data and in section \ref{sect_discus} we present the final
discussion of our results.

\section{interference mechanism of magnetoresistance}
\label{sect_gen}

It is convenient to describe hopping conductivity in terms of
random resistance network first proposed by Miller and Abrahams \cite{Miller-Abr}.
In terms of this model any pair of impurities (denoting as $1$ and
$2$) corresponds to a resistor with resistivity
\begin{equation} \label{Rij}
R_{12} \propto \Gamma_{12}^{-1}, \quad \Gamma_{12} \propto
\left< \left|\widetilde{I}_{12}\right|^2 n_1 (1-n_2)  N_{ph}
\right>.
\end{equation}
Here $\Gamma_{12}$ is the hopping rate, $n_1$ and $n_2$ are the
occupation numbers of the first and the second impurity states,
correspondingly. $\widetilde{I}_{12} \propto \exp(-r_{12}/a)$ is
the energy overlap integral between states $1$ and $2$, $r_{12}$
is the distance between impurities and $a$ is the localization
length. $N_{ph}$ is the probability to find a phonon for a hop
(one can take $N_{ph}=1$ for hops with emission of a phonon).
Finally, angle brackets mean time averaging.

The terms $n_1$, $(1-n_2)$ and $N_{ph}$ specify the dependence of
resistor $R_{12}$ on impurity energies. We will not discuss this
part of (\ref{Rij}). What we are interested in is the overlap
integral $\widetilde{I}_{12}$. When there are no impurities beside
$1$ and $2$, the overlap integral can be easily evaluated
\cite{Book}
\begin{equation} \label{Iij0}
\widetilde{I}_{12} = I_0 \left( \frac{r_{12}}{a} \right)^{\beta}
\exp(-r_{12}/a).
\end{equation}
Here $I_0$ is the constant of the order of the Bohr energy, $\beta$
is the pre-exponential factor that is determined by the impurity
type. For shallow impurities in the bulk system (or in the
$\delta$-doped layer) $\beta = 1$. For non-Coulomb impurities
(like the deep impurities or doubly-occupied $D^-$ or $A^+$
centers) $\beta=-1$. In two-dimensional systems $\beta$ is equal
to $1/2$ and $-1/2$ correspondingly for Coulomb and non-Coulomb
impurities.

When there are intermediate scattering impurities between the hopping ones, the overlap
integral becomes a sum of the different tunneling paths and
expression (\ref{Iij0}) stands only for one of them (the one
without scattering). The total overlap integral between impurities
$1$ and $2$ can be expressed as
\begin{equation} \label{Iij}
\widetilde{I}_{12} = I_{12} + \sum_i \frac{I_{1i}I_{i2}}{E_i} +
\sum_{ij} \frac{I_{1i}I_{ij}I_{j2}}{E_iE_j} + ...
\end{equation}
Here  $I_{ij}$ are defined by (\ref{Iij0}). $E_i$ are the energy
positions of scattering impurities (with respect to the Fermi
level). We assumed that the corresponding energies are much larger
then the energies of hopping impurities $1$ and $2$. Each summand
in (\ref{Iij}) corresponds to an amplitude of some tunneling path.
Let us note that any tunneling path amplitude is  $ \propto
\exp(-l/a)$ where $l$ is the length of the corresponding hopping
segment. So the paths with large $l$ can be neglected.
Accordingly, we neglect all paths with backscattering and also
include in our consideration only scattering impurities in some
thin strip (in real space) around the line, connecting impurities
$1$ and $2$. The width of this strip is estimated in \cite{SS} as
$\sqrt{r_{12}a}$. However we will show that in the case of long
hops it is sufficient to consider a thinner strip. So we have a
finite number of paths $2^N$ where $N$ is the number of scatterers
in the strip.

Let us also write the expression for the simplest (but important)
case of one scattering impurity $s$. In this case
\begin{equation}\label{G1}
\Gamma_{12} \propto \left|J_1+J_2\right|^2, \quad J_1 = I_{12},
\quad J_2 = \frac{I_{1s}I_{s2}}{E_s}.
\end{equation}

Note that the expression (\ref{G1}) and the upper expressions do not
take in account the electron spin. Consequently up to this moment we
discussed the model of spinless electrons. As we have already
mentioned, this model can be justified when electron spins on
scattering impurities are frozen (for details see \cite{SS}). But
now let us assume that the intermediate impurity is occupied and
its electron spin is free.

\begin{figure}[htbp]
    \centering
        \includegraphics[width=0.5\textwidth]{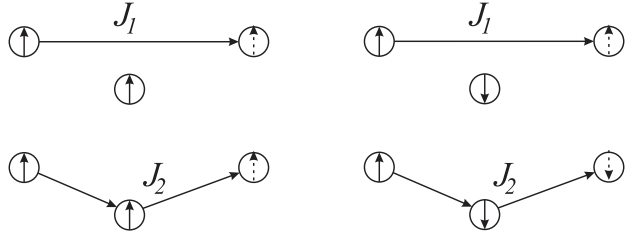}
        \caption{Final results of direct and scattered tunneling path in different spin configuration.}
    \label{fig:frezfree}
\end{figure}

The scattering on the occupied impurity is equivalent to the
cotunneling of the initial electron to the scatterer and the
electron from the scatterer to the destination impurity (figure
\ref{fig:frezfree}). One can see that when spins of initial
electron and  electron on the scatterer have the same
directions, the final result of direct tunneling $J_1$ and the
tunneling with scattering $J_2$ are the same. However, when the
directions of these spins are antiparallel, the final results of
the trajectories $J_1$ and $J_2$ are different. So in this case
the tunneling paths do not interfere.

As long as the spin on the scattering impurity is free, there is
always a probability that the two spins have the same direction
and a probability that they have antiparallel directions. Note
that the expression (\ref{Rij}) for $\Gamma_{ij}$ contains time
averaging. So for the free spin on the intermediate impurity one
obtains
\begin{equation} \label{G2}
\Gamma_{12} \propto P_{\uparrow\uparrow}\left|J_1 + J_2\right|^2 +
\left(1 - P_{\uparrow\uparrow} \right)\left( |J_1|^2 +
|J_2|^2\right).
\end{equation}
Here $P_{\uparrow\uparrow}$ is the probability for  two spins
to have the same direction. Without the magnetic field
$P_{\uparrow\uparrow}=1/2$ because both spin directions are
equally probable. The external magnetic field aligns the spins leading to an increase of
$P_{\uparrow\uparrow}$. The expression for probability
$P_{\uparrow\uparrow}$ in a magnetic field is as follows
\begin{equation} \label{PH}
P_{\uparrow\uparrow} = \frac{\cosh(2\mu H/T)}{2\cosh^2(\mu H/T)} =
\frac{1}{2} + \frac{1}{2}\left(\frac{\mu H}{T}\right)^2 + O(H^4).
\end{equation}
Here $\mu$ is the magnetic moment of a localized electron, $H$ is
the external magnetic field and $T$ is temperature expressed in
energy units.

In the expression (\ref{G2}) we have implicitly assumed that the
spin states change with time faster than the hopping events occur
or at least faster than Miller-Abrahams network is stabilized. To
explain this assumption we note that the spin states at least can
not change significantly slower then the hopping rate because the
hopping in the antiparallel spin combination can exchange the
spins. Also, other mechanisms of spin diffusion or
relaxation can exist, for example, related to the exchange of spin
directions between neighbor impurities with free spin. We also
note that our expression (\ref{G2}) agrees with expression (14)
from \cite{Spivak} (that describes spin alignment
magnetoresistance).

 There is another assumption that we will
rely upon. Namely, we will make use of the fact that the frozen
spins form the so called Bhatt-Lee phase \cite{BhattLee}. In this
state the frozen spins are in a singlet state and thus have no
preferred direction. So the frozen spins scatter all tunneling
electrons in the same way (independently of their spin direction).
If by some reason the frozen spins form spin glass state, our
considerations can not be applied.

One can see that there is a dependence of $\Gamma_{12}$ on
magnetic field even if the intermediate impurity has a free spin.
First, in magnetic field a phase difference between
tunneling paths exist. It can be expressed as $J_2 \rightarrow
J_2e^{i\varphi}$, $\varphi = HS/\Phi_0$, where $S$ is the area
between the tunneling paths, $\Phi_0$ is the magnetic flux
quantum. Secondly, the probability $P_{\uparrow\uparrow}$ depends
on the magnetic field (see Eq. (\ref{PH})). According to the
logarithmic averaging procedure (\cite{SS}), one can write the
following expression for the contribution of scattering on free
spins states to the magnetoresistance:
\begin{equation} \label{MR0free}
\ln \frac{R(H)}{R(0)} = - \left< \ln
\frac{P_{\uparrow\uparrow}\left|J_1 + J_2 e^{i\varphi}\right|^2 +
\left(1 - P_{\uparrow\uparrow} \right)\left( |J_1|^2 +
|J_2|^2\right)}{\left|J_1 + J_2\right|^2/2 + \left( |J_1|^2 +
|J_2|^2\right)/2} \right>_{free}.
\end{equation}
Here the angle brackets with index "free" mean the averaging over
critical resistors with free spin on the scatterer (note that in
short hop limit we have only one scatterer). There is also a
contribution from resistors with unoccupied intermediate impurity
and from resistors involving a scattering impurity with frozen
electron spin. These contributions have a similar form:
\begin{equation}\label{MR0fro}
\ln \frac{R(H)}{R(0)} = - \left< \ln \frac{\left|J_1 + J_2
e^{i\varphi}\right|^2 }{\left|J_1 + J_2\right|^2} \right>_{frozen}.
\end{equation}
Note that angle brackets with index "frozen" mean not only the
averaging over scatterers with an electron with frozen spin but also
over the unoccupied scatterers. The contribution of these two
scatterer types have the same form although the effects of these
scatter types are different due to the sign of energy $E_s$.

Now we generalize the discussed approach for the situation when  a
resistor includes $N$ scattering impurities and $M$ of them are
occupied by  electrons with free spins. When a resistor contains
one free spin, there are four possible initial spin configurations
(corresponding to up and down projection of initial spin and free
spin on the scatterer). Similarly, $M$ free spins in a resistor
lead to $2^{M+1}$ possible initial spin configurations. Also, in a
resistor with one free spin one has no more then two possible tunneling results
(final states of the tunneling) for any initial configuration
 --- with and without
exchange of  spin projections. In our resistor one has no more
than $2^M$ tunneling results for any spin configuration. These
results can be described by the combination of free spins that
changes the spin projection during the process of hopping. Note
that some results are impossible in specific spin configurations
(for example for one intermediate free spin there is only one
tunneling result for spin configurations with parallel spin
projections).

Accordingly, for any spin configuration $2^N$ tunneling amplitudes
can be separated into $2^M$ groups corresponding to different
tunneling results (some groups can contain no tunneling amplitudes
in some initial spin configurations). The tunneling paths that are
inside one group interfere (because they lead to the same final state
in the given initial configuration of free spins). Other tunneling
paths do not interfere. One should average $\Gamma_{12}$ over
initial configurations:
\begin{equation} \label{MR_Many_G}
\Gamma_{12} \propto  \displaystyle \sum_{c=1}^{2^{M+1}} P_c(H)
\sum_r \left|\sum_{k(c,r)} J_k e^{i\varphi_k}\right|^2.
\end{equation}
Here index $c$ enumerates possible spin configurations, $P_c(H)$
is the probability of a given spin configuration in the magnetic
field to exist. Index $r$ enumerates the tunneling results. $J_k$
are the amplitudes of the tunneling paths and $\varphi_k$ are the
corresponding phases. The sum over $k$ is taken only over those
tunneling paths that lead to result $r$ in the spin configuration
$c$. The probability $P_{c}(H)$ can be expressed as
\begin{equation}\label{Pc(H)}
P_c(H) = \frac{\exp\left[(2N_{up}(c) - M - 1)\mu H/T
\right]}{2^{M+1}\cosh^{M+1}(\mu H/T)},
\end{equation}
where $N_{up}(c)$ is the number of free electron spins (including the spin of tunneling electron) that are
aligned along the magnetic field in the configuration $c$.

The corresponding expression for magnetoresistance is
\begin{equation} \label{MR_Many_M}
\frac{R(H)}{R(0)} = - \left\langle \ln \frac{\displaystyle
\sum_{c=1}^{2^{M+1}} P_c(H) \sum_r \left|\sum_{k(c,r)} J_k
e^{i\varphi_k}\right|^2}{\displaystyle \sum_{c=1}^{2^{M+1}} P_c(0)
\sum_r \left|\sum_{k(c,r)} J_k\right|^2} \right\rangle.
\end{equation}
Here the angle brackets mean averaging over critical resistors. We
note that number $M$, possible spin configurations and tunneling
results can be different for different critical resistors.

 When
there are no resistors with two or more scatterers, the expression
(\ref{MR_Many_M}) is reduced to (\ref{MR0free}) and
(\ref{MR0fro}). When all electron spins are frozen, it is reduced
to the conventional expression for magnetoresistance from
\cite{SS} (that corresponds to $\Gamma_{12}\propto
|\widetilde{I}_{12}|^2$ with $\widetilde{I}_{12}$ defined in
(\ref{Iij})).

The expression (\ref{MR_Many_M}) is rather complex. However it
allows numerical computation of magnetoresistance for a relatively
large number of scatterers. The results of such computations
will be presented in the section \ref{sect_lonhhops2} of this study
 along with analytical analysis of the magnetoresistance.

\section{Short hops}
\label{sect_shorthops}

 In the case of short hops one can
neglect resistors with more than one scatterer. Hopping events
with no scatterer do not contribute to interference
magnetoresistance so that one can consider only resistors with one
scatterer. For those resistors it is possible to use expressions
(\ref{MR0free}) and (\ref{MR0fro}).

The magnetoresistance corresponding to short hop limit with no
free electron spins (expression (\ref{MR0fro})) is known from
\cite{Shirmacher,Raikh}. It can be shown that the averaging in
(\ref{MR0fro}) yields different results for the two areas. In the
first area, $A_1$, the denominator is not very small
$|J_1+J_2|>\varphi J_1$. In this area one can expand the
logarithm. It leads to a quadratic magnetoresistance. In the
second area, $A_2$, where $|J_1+J_2|< \varphi J_1$, the logarithm
is large and can not be expanded. This area leads to the negative
linear magnetoresistance that dominates in weak fields. For short
hops the expression for this magnetoresistance is (see
\cite{Shirmacher,Raikh})
\begin{equation} \label{MR1fro}
\ln \frac{R(H)}{R(0)} \propto -  r_h^{2+d/2} H,
\end{equation}
where $r_h$ is the mean hopping distance and $d$ is the system
dimensionality. The dependence on $r_h$ controls the temperature
dependence of the magnetoresistance as $r_h \propto T^{-1/(d+1)}$
for the Mott-like hopping and $r_h \propto T^{-1/2}$ for
Efros-Shklovskii hopping over Coulomb gap states.

For the free electron spins the area $A_2$ does not exist.
Actually, the denominator in (\ref{MR0free}) is not less than
$3|J_1|^2/4$. The denominator in (\ref{MR0fro}) for resistors with
unoccupied scatterers (that has constructive interference) is
larger then $|J_1|^2$. Hence in weak magnetic field the logarithm can always
be expanded. This leads to the following expression for
the weak field magnetoresistance expansion:
\begin{equation} \label{MR1free}
\ln \frac{R(H)}{R(0)} \approx - \left\langle  \frac{
 (\frac{\mu_bgH}{T})^2 J_1J_2 - \frac{1}{2}J_1J_2 \varphi^2}
{\frac{1}{2} (J_1+J_2)^2 + \frac{1}{2} (J_1^2+J_2^2)}
\right\rangle_{occupied} + \left\langle  \frac{
 J_1J_2 \varphi^2}
{ (J_1+J_2)^2} \right\rangle_{unoccupied}.
\end{equation}
Here angle brackets with index "occupied" mean averaging over
resistors with occupied scattering impurity (in this expression we
consider all such impurities to have a free electron spin). Note
that for this impurities $J_1J_2$ is negative. Angle brackets with
index "unoccupied" corresponds to averaging over resistors with
free intermediate impurity. For those resistors $J_1J_2$ is positive.

There are three contributions to magnetoresistance \textit{due to
interference} for free electron spins. First, there is one
negative contribution. It comes from  magnetic field suppression
of destructive interference for the resistors with occupied
scatterers. However, in contrast to the situation when electron spins
are frozen, it is quadratic (not linear) in terms of the magnetic
field and does not automatically dominate over other terms in a
weak field limit. Also, it does not automatically dominate over
the wave shrinkage magnetoresistance even for weak fields.

The second term appears from the suppression of interference in
resistors with unoccupied scatterers and constructive
interference. It will dominate the first term at least for
semiconductors with large compensation $K$ ($1-K \ll 1$), as the
number of unoccupied scatterers in this semiconductors is larger
than the number of occupied ones.

The third term in (\ref{MR1free})  is also
positive and quadratic in terms of magnetic field
\begin{equation} \label{3rd_term}
- \left\langle  \frac{
 (\frac{\mu_bgH}{T})^2 J_1J_2}
{\frac{1}{2} (J_1+J_2)^2 + \frac{1}{2} (J_1^2+J_2^2)}
\right\rangle_{occupied} \propto (n
r_h^{(d+1)/2}a^{(d-1)/2})\left(\frac{\mu_bgH}{T}\right)^2.
\end{equation}
Here $n$ is the impurity concentration, $d$ is the system
dimensionality. On the right-hand side of the equation we used the
following approximation. We considered $J_1$ and $J_2$ to be of
the same order (it is usually correct if the scatterer lies in the
thin area between the hopping impurities \cite{SS}).  $(n
r_h^{(d+1)/2}a^{(d-1)/2})$ is the probability to find a scatterer
in this area.

The term (\ref{3rd_term}) is related to the dependence of the
probability $P_{\uparrow\uparrow}$ on the magnetic field. It is
the spin-ordering magnetoresistance first discussed in
\cite{Spivak}. One can see that this term has very strong
temperature dependence. For example, for a 2D system with
hopping over the Coulomb gap states, this term depends on
the temperature as $T^{-11/4}$. To compare, the temperature dependence
of the wave-shrinkage magnetoresistance in this case is $\propto
T^{-3/2}$. Therefore at low temperature the spin-alignment
magnetoresistance should become stronger than the wave-shrinkage
one.

However, at low temperatures the system should correspond to the
case of long hops  (rather then of the short hops) and  at
some temperature electron spins should be controlled by the
exchange interaction. However (as we show in section
\ref{sect_lonhhops2}) the spin alignment magnetoresistance remains
in the long hop limit and  has a strong temperature dependence.

Finally, one can consider the situation when some spins are frozen
and some spins are free, that is $0<P_{free}<1$. The short hops
theory of magnetoresistance can be easily generalized for this
case. Naturally, each resistor contains either a free spin or a
frozen one (if it involves an occupied scatterer). The general
expression for magnetoresistance contains averaging over critical
resistors. Accordingly the following equation can be used for shot hops
$$
\ln\frac{R(H)}{R(0)} = P_{free} MR_{free} + (1-P_{free})
MR_{frozen},
$$
where $MR_{free}$ is given by the expression (\ref{MR1free}) and
$MR_{frozen}$ is given by the expression (\ref{MR1fro}).
Naturally, when there is a significant number of frozen spins in
the system ($1-P_{free}$ is not very small) the weak field
magnetoresistance will be controlled by $MR_{frozen}$ and will be
negative and linear in terms of  magnetic field. However, if one
considers the quadratic term of the magnetoresistance expansion,
then the term $MR_{free}$ can also become important.

\section{Long hops. Statistics of tunneling paths}
\label{sect_longhops1}

We address now the case of long hops, i.e. the situation when
there is a large number of scatterers in a hopping process. As we
have mentioned before, in this case there is exponentially large
number of tunneling paths $2^N$ (even without backscattering),
where $N$ is the number of intermediate impurities in the
resistor. When there are no free spins, all paths interfere. The
situation with free spins is more complex as it was discussed in
section \ref{sect_gen}.

In our opinion, the role of free spins in this case is  even more
important than it is for short hops. Indeed, for sufficiently long
hops free spins are always present in the area between hopping
sites.  It is known that with decreasing temperature the spins of
electrons that are localized on the impurities form the so-called Bhatt-Lee
phase \cite{BhattLee}. One of the properties of this phase is
a relatively large number of free spins. The simplest estimate of
the number of free spins is based on the nearest-available-neighbor
approach \cite{NAN}. It leads to the logarithmic decrease of the free
spin concentration with decreasing temperature. The hopping length
grows with temperature according to a power law. So at low
temperatures the number of scatterers with free spins in a hopping
process is large.

In \cite{SS} the following arguments for the absence of
interference magnetoresistance in the case of free electron spins
has been proposed. It has been stated that in this case there are many
tunneling paths that do not interfere with each other, making the
strong destructive interference (that is responsible for linear
negative magnetoresistance)  very unlikely.

 However,  at
sufficiently low temperatures each critical resistor has free
spins and, therefore, there is always a large number of tunneling
paths that do not interfere (there are $2^M$ possible tunneling
results, where $M$ is the number of scatterers with free spin in
the hopping act). Consequently the quantitative estimate of the effect of
the free spins on interference magnetoresistance required. The present study provides
such an estimate.

The role of interference effects in the magnetoresistance in the
limit of long hops has been discussed (although without free spin consideration)
in a number of publications \cite{PRL_perc,
Medina,PRL_an,an,Spin-Orb}. However all these reports imply two
important model assumptions which we do not believe to be obvious.

The first assumption is placing of scattering impurities in the
nodes of some lattice. Usually one places scatterers in the nodes
of a finite square lattice. The hopping impurities are on the
diagonally opposite corners of the lattice. Then, one considers
the shortest hopping paths that are along the lattice bonds
(there are many such paths with equal length). In the present study we
do not discuss the applicability of this assumption. However we do
not use it.

The second assumption implies a neglect the variance of scatterer energies.
In \cite{SS, Medina, PRL_an,Spin-Orb} the authors
consider scatterers with two possible energies $W$ and $-W$, so
that the variance of the absolute value of the scatterer energy
was neglected. In \cite{PRL_perc, an} the authors consider the
flat distribution of scatterer energies at least as one of the
possible models. However, then they assumed that the
distribution of tunneling path amplitudes is the Gaussian one with
the relative variance of the order of unity.

We argue that the second assumption is not correct for realistic
semiconductors. Actually, the scatterer energies in semiconductors
have flat distribution and this fact leads to the log-normal
distribution of tunneling path amplitudes which is significantly
different from a normal one. This distribution plays a crucial
role in our theory.

Let us consider some tunneling path. In accordance to the
expression (\ref{Iij}), its amplitude can be given as
\begin{equation} \label{J1}
J_i = I_{0,k_1}\prod_{k(i)}^{N_i} \frac{I_{k,k+1}}{E_k}.
\end{equation}
Here $J_i$ is the amplitude of i-th tunneling path. The index
$k(i)$ enumerates the scattering impurities that participate in
the tunneling path $i$. $N_i$ is a number of these impurities.
Usually $N_i \sim N/2$, where $N$ is the total number of
scatterers in the resistor. $k_1$ is the first of the impurities
of this path.
 Index $0$ ($k=0$) stands for the
initial hopping impurity. The last index $k=N_i+1$ corresponds to
the destination impurity of the hop.

Let us take a logarithm of the absolute value of $J_i$ and write
explicitly the term that is related to the hopping distance.
\begin{equation} \label{J1}
\ln \left|\frac{J_i}{I_0}\right| = -r_h/a + \sum_{k(i)} \mu_k^i, \quad
\mu_k^i = -\frac{\Delta r_k}{a} +  \ln \frac{I_0}{|E_k|} + \beta \ln
\frac{r_{k-1,k}r_{k,k+1}}{r_{k-1,k+1}a}.
\end{equation}
Here $\Delta r_k$ is the additional distance an electron should
tunnel due to scattering on the impurity $k$, $\Delta r_k =
r_{k-1,k}+r_{k,k+1} - r_{k-1,k+1}$.  $r_{i,j}$ is the distance
between scattering impurities $i$ and $j$, it may occur that some
of this indices ($i$ and $j$) are equal to zero or to $N_i+1$
which means the starting impurity of the hop and the destination
impurity, correspondingly. $\beta$ is the pre-exponential factor
in the overlap integral (see (\ref{Iij0})).

We consider the scattering impurities to be situated in a thin strip
to prevent the additional distance to be too large, $\Delta r_k
\leq a$. Also, the typical values of scatterer energies are of the
order of the Bohr energy, so $I_0/|E_k| \sim 1$ (the sign of $E_k$
is random). Finally, the third term is of the order of $\ln
r_{sc}/a$ where $r_{sc}$ is the characteristic distance between
scatterers in the resistor. In real situations this logarithm is
not very large.
 Accordingly, we can consider $\mu_k^i$ as a random value
with expectation and variance of the order of unity.

Sometimes it is more useful to sum not over scatterers
participating in the path $i$, but over all scatterers in the
resistor. Therefore we introduce the modified values
$\widetilde{\mu}_j^i$ that are equal to $\mu_j^i$ if scatterer $j$
participates in the given path and equals to 0 otherwise.
\begin{equation}
\widetilde{\mu}_{j}^i = \left\{
\begin{array}{ll}
\mu_j^i, \quad & j \in {\rm path} \,\, i\\
0 & {\rm otherwise}
\end{array} \right.
\end{equation}
Again, we consider $\widetilde{\mu}_j^i$ as random quantities with
math expectation $E_\mu$ and variance $D_\mu$ of the order of
unity.

If one takes $\widetilde{\mu}_j^i$ as independent quantities (this
assumption will be discussed later), one gets the normal
distribution for $\ln |J_i/I_0|$ (in accordance with the central
limit theorem). The math expectation of $\ln |J_i/I_0|$ is $-
r_h/a+NE_\mu$ and its variance is $N D_\mu$.

Hence the quantity $|J_i|$ has a log-normal distribution law
\begin{equation} \label{distribJi}
f(|J_i|) = \frac{1}{|J_i| \sqrt{2\pi N D_{\mu}}} \exp \left[ -
\frac{(\ln (|J_i|/I_0) + r_h/a - NE_\mu)^2}{2(ND_\mu)^2} \right], \quad
|J_i|>0.
\end{equation}
The math expectation $E(|J_i|)$ and variance $D(|J_i|)$ are expressed
as follows:
\begin{equation}\label{E(J)}
E(|J_i|) = I_0 \exp \left( -\frac{r_h}{a} + N E_\mu + \frac{ND_\mu}{2}
\right);
\end{equation}
\begin{equation}\label{D(J)}
D(|J_i|) = I_0^2 \exp \left( -\frac{2r_h}{a} + 2 N E_\mu +  2 ND_\mu
\right).
\end{equation}

This distribution law is significantly different from the one that
can be obtained from normal distribution of $J_i$. Actually, it is
the exponentially broad distribution. The important feature of the
law (\ref{distribJi}) is that its math expectation (\ref{E(J)}) is
exponentially larger than the value corresponding to the maximum
of distribution density $I_0 \exp \left( - r_h/a + N E_\mu\right)$.
It means that a sum of many tunneling path amplitudes (for example
$\widetilde{I}_{12}$) should be dominated by exponentially small
number of tunneling paths. In the next section we use this fact to
derive a theory of interference magnetoresistance in the limit of
long hops.

However, before we will discuss the magnetoresistance, let us consider
 the following moment. We derived the log-normal distribution
(\ref{distribJi}) from the assumption that the quantities
$\widetilde{\mu}_j^i$ where independent. Strictly speaking, this
assumption is incorrect. The same impurities that form the
interference pattern of the resistor participate in all
tunneling paths (although in different combinations). Consequently we
believe that it is reasonable to check the log-normal distribution
by numerical calculations.

\begin{figure}[htbp]
    \centering
        \includegraphics[width=0.8\textwidth]{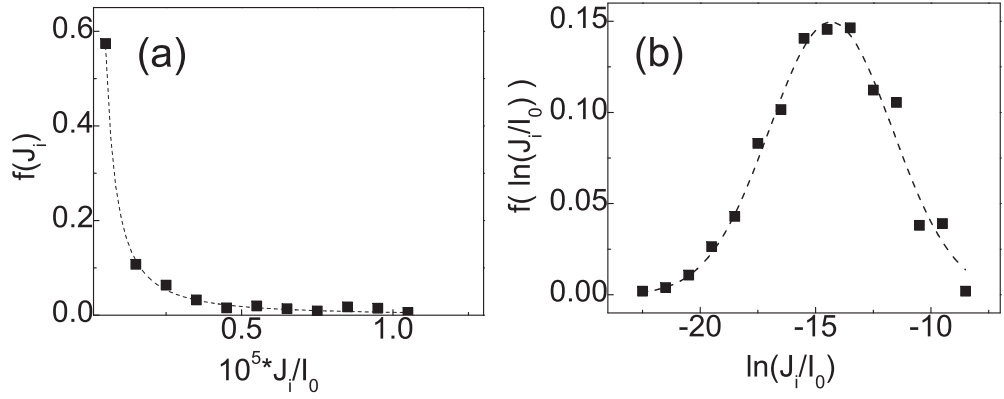}
        \caption{ Distribution functions of $J_i$ $(a)$ and $\ln
        J_i/I_0$ $(b)$ calculated for a tunneling path amplitudes in a
        single resistor compared to the log-normal distribution.
         }
    \label{fig:fJi}
\end{figure}

Let us take the resistor with 10 scattering impurities. This
resistor has $2^{10}=1024$ tunneling paths, which is enough to reach
the reliable statistics. The corresponding results for tunneling
path amplitude distribution are shown on fig. \ref{fig:fJi}. One
can see that they are in a good agreement with log-normal law.

Another important property of the obtained tunneling path
amplitude distribution is the fact that the sum of all tunneling
paths is dominated by a small number of the summands. Let us in
addition verify this result. In order to do this, we have to formalize the
concept "dominate". We do this in the following way.

Consider the sum of all absolute values of tunneling path amplitudes
$\sum_i |J_i|$. This sum has $2^N$ summands. Let us now take some
number $N_{sign}$ of its largest summands, so that
\begin{equation}\label{Nsign}
\sum_k^{N_{sign}} |J_k| \ge 0.6 \sum_i^{2^N} |J_i|.
\end{equation}
The left hand part of the inequality is the sum of $N_{sign}$
largest $|J_i|$. The $N_{sign}$ is considered to be the smallest
number to satisfy inequality (\ref{Nsign}). We will specify
$N_{sign}$ tunneling paths with largest absolute values of
amplitude as the significant paths.

\begin{figure}[htbp]
    \centering
        \includegraphics[width=0.8\textwidth]{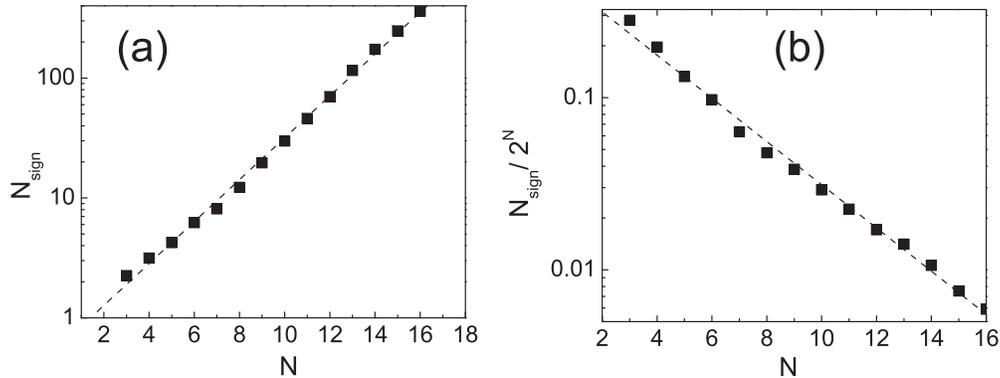}
        \caption{ (a) --- dependence of $N_{sign}$ on the scatterer number $N$.
        (b) ---  relative part of the significant paths
        $N_{sign}/2^N$ versus scatterers number.
         }
    \label{fig:Nsign}
\end{figure}

If the most of the tunneling paths are important within the sum
$\sum_i |J_i|$, one should get $N_{sign} \sim 0.5\cdot 2^N$.
However if our assumption is correct and $\sum_i |J_i|$ is
controlled by a small number of its summands, we should have
$N_{sign} \ll 2^N$. The corresponding numerical results are shown on
figure (\ref{fig:Nsign}). One can see that, although the number
$N_{sign}$ exponentially grows with $N$,
  the relative part of significant paths exponentially
decreases with $N$, demonstrating the validity of our assumption
that the sum of contributions of different paths is controlled by a small number of summands.
 In the next section we will exploit it do
derive the expression for magnetoresistance in the long hop limit.

\section{Magnetoresistance in the limit of long hops}
\label{sect_lonhhops2}

In the spinless electron model the interference magnetoresistance
is controlled by the overlap integral $\widetilde{I}_{12}$, which
is the sum of many tunneling path amplitudes $J_i$. We have shown
that this sum is controlled  by a relatively small number of the
largest $J_i$. Now we consider the question which of $J_i$
occurs to be large. $J_i$ is proportional to the product
$\prod_{j}\exp({\mu}_{j}^i)$ for  impurities $j$ participating in
path $i$. The values ${\mu}_j^i$ control whether the path
$i$ gives larger or smaller amplitude  with an inclusion of the
impurity $j$.

Strictly speaking, ${\mu}_j^i$ is controlled not only by the
properties of impurity $j$ but also by other impurities
participating in path $i$. However, we can estimate $\mu_j^i$ for
any scatterer $j$ and a characteristic tunneling path.
\begin{equation}\label{muj}
\overline{\mu_j} = -\frac{y_j^2}{2r_{sc}a} + \ln \frac{I_0}{|E_j|} + \beta\ln
\frac{r_{sc}}{2a}.
\end{equation}
Here $y_j$ is the distance between impurity $j$ and the  line connecting the starting and the final impurities of the hop,
 $r_{sc}$ is the characteristic distance between
the scatterers in the tunneling path.  $r_{sc}$ will be estimated later in our paper.

There are three opportunities for $\overline{\mu_j}$. First it may
occur that $\overline{\mu_j}$ is negative and has a large modulus,
so $\exp \overline{\mu_j} \ll 1$. The corresponding impurities
usually decrease the tunneling amplitude. The inclusion of a large
number of  such impurities makes the tunneling path amplitude
exponentially small, so that this path will not be significant. Let us
call such impurities with small $\overline{\mu_j}$ as the irrelevant
scatterers. Actually all impurities that are far from the line
connecting the hopping impurities are irrelevant.

In contrast, it can occur that the absolute value $|E_j|$ is much
smaller then $I_0$  and $\exp \overline{\mu}_j \gg 1$. Inclusion
of such impurities makes the path amplitude larger. If some
tunneling path does not include many such impurities, its
amplitude becomes exponentially smaller than amplitudes of other
paths (which include the corresponding impurities) and this path
can not be significant. So the impurities mentioned above should
exist in most significant paths. We call these impurities as the
backbone impurities.

Finally, there are impurities with $\exp \overline{\mu}_j \sim 1$.
They can be included and can be excluded from a significant path.
We call these impurities as the interference impurities.

\begin{figure}[htbp]
    \centering
        \includegraphics[width=0.5\textwidth]{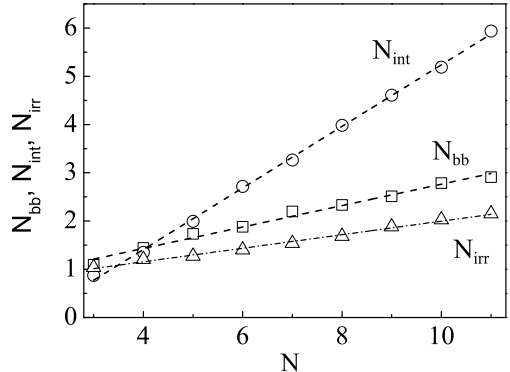}
        \caption{ Dependance of mean numbers of the different impurity types ($N_{bb}$, $N_{int}$ and $N_{irr}$) on
        the total number of scatterers $N$. Each dot is averaged over 100 realizations.
         }
    \label{fig:carcass}
\end{figure}

\begin{figure}[htbp]
    \centering
        \includegraphics[width=0.95\textwidth]{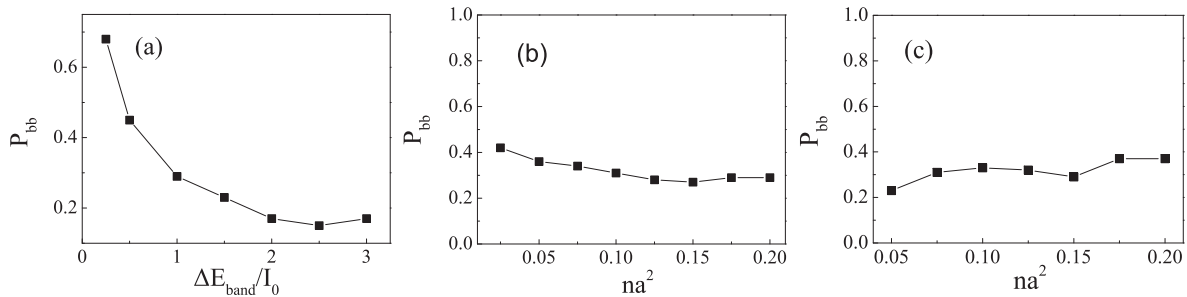}
        \caption{ Dependance of the relative number of backbone impurities $P_{bb}$ on the system parameters.
        (a)---  dependance on the width of impurity band $\Delta E_{band}$ with $na^2=0.2$ and pre-exponent factor $\beta=1$, (b) ---  dependance on
        the impurity concentration with $\Delta E_{band} = I_0$ and $\beta = 1$, (c) ---  dependance on
        the impurity concentration with $\Delta E_{band} = I_0$ and $\beta = -1$.
         }
    \label{fig:carcass2}
\end{figure}

Strictly speaking, the separation of impurities to irrelevant,
backbone and interference ones is valid only for the exponentially
broad distribution of scattering impurity energies (in this case
$\ln I_0/E_j$ is the leading term in (\ref{muj}) and can be
large).
 However, even if the distribution is not exponentially broad,
there are some impurities with small value of $|E_j|$ that are
good candidates to be the the backbone impurities. Let us try to
find such impurities in numerical computations. To perform the
computations, we consider impurity to be a backbone one when 90\%
of the significant paths include this impurity. Similarly, the
irrelevant impurities are the ones that are not included in 90\%
of significant tunneling paths.

Figure \ref{fig:carcass} shows the averaged results for resistors
with different length and with a constant width of the area
occupied by the scatterers. One can see that the number of
backbone impurities $N_{bb}$ depends linearly on the scatterer
number $N$. So we can estimate the relative number of backbone
impurities in the long resistor limit. The same procedure can be
followed for  numbers of interference and irrelevant
impurities $N_{int}$ and $N_{irr}$. Note that the relative parts
of different types of impurities can depend on the width of the
area exploited for calculations. For  very wide area most
impurities will be irrelevant. So we redefine the relative number
of backbone impurities to make it model-independent
\begin{equation}
P_{bb} = \lim_{N \rightarrow \infty} \frac{N_{bb}}{N_{bb} + N_{int}}.
\end{equation}
$P_{bb}$ is the relative part of backbone impurities with
irrelevant impurities excluded. It does not depend on the the
width of the model area if it is large enough. Fig.
\ref{fig:carcass2} shows our computation of the dependence of
$P_{bb}$ on the impurity band width $\Delta E_{band}$,
concentration $n$ and pre-exponent factor $\beta$. One can see that
$P_{bb}$ increases with the decrease of $\Delta E_{band}$. For
large band width $P_{bb}$ should tend to some finite value as the
impurities with large energy are always  irrelevant scatterers.
$P_{bb}$ weakly depend on concentration and pre-exponent factor at
least in observed range of parameters.

Now let us make use of the existence of backbone impurities to
calculate the magnetoresistance in the long hops regime. In the
present study we restrict ourselves with the simplest model of
independent links. This model assumes that any two interference
impurities are always separated with a backbone one. Therefore this model
should be valid for $P_{bb} > 0.5$. However the results of our
computations agrees with this model even for smaller $P_{bb}$.

\begin{figure}[htbp]
    \centering
        \includegraphics[width=0.6\textwidth]{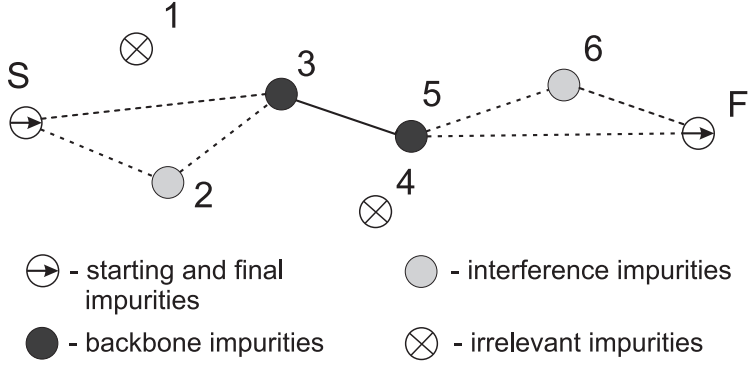}
        \caption{ The example of the hopping resistor with six scatterers. The scattering impurities are numerated while the starting
        and final impurities in this scatterer are marked with indexes $S$ and $F$ correspondingly.
         }
    \label{fig:pic1}
\end{figure}

Let us illustrate our considerations with the following model
resistor (see fig. \ref{fig:pic1}). We consider, first, the
spinless electron model. This resistor contains two backbone
impurities, two interference impurities and two irrelevant ones.
So the total number of scatterers is $N=6$, the number of
tunneling paths is $2^N = 64$. However, in our model we assume
that one can restrict himself with consideration of significant
paths ---  ones that contain all backbone scatterers and bypass all
irrelevant impurities. Therefore instead of 64 tunneling paths one can
consider only four ones: $(S \rightarrow 3 \rightarrow5
\rightarrow F)$, $(S \rightarrow 2 \rightarrow 3 \rightarrow5
\rightarrow F)$, $(S \rightarrow 3 \rightarrow 5  \rightarrow
6\rightarrow F)$ and $(S \rightarrow 2 \rightarrow3 \rightarrow5
\rightarrow 6 \rightarrow F)$.

The amplitudes of these paths are expressed by Eq. (\ref{J1}). One can
see that all  amplitudes of these paths contain the term
$I_{35}/E_3E_5$. It corresponds to  backbone impurities. Then, all
 paths contain connections between the backbone parts. For
example, impurities $S$ and $3$ can be connected with either
$I_{S3}$ or $I_{S3}I_{23}/E_{2}$. One can show that within our
approximations these connections are independent, i.e. the net
tunneling amplitude $J = \sum_i J_i$ can be expanded as
\begin{equation}
\label{chain1} J = \left(I_{S3} + \frac{I_{S2}I_{23}}{E_2}\right) \frac{I_{35}}{E_3E_5}\left(I_{5F} + \frac{I_{56}I_{6F}}{E_6}\right).
\end{equation}

The backbone impurities effectively separate the resistor into
independent parts --- the interference links. The amplitude of each
link can be described in the same way as a tunneling amplitude for
the resistor with one scatterer. The net tunneling amplitude of
the resistor in the independent links model is the
product of terms corresponding to all links and additional
terms corresponding to backbone impurities.
\begin{equation} \label{chain2}
J = \prod_{backbone}\frac{I_{kj}}{E_j}\cdot \prod_{interference} \left(I_{i1,i3} + \frac{I_{i1,i2}I_{i2,i3}}{E_{i2}} \right).
\end{equation}
Here the first term is the product of $1/E_j$ for all backbone
impurities $j$ and the product of all the overlap integrals $I_{kj}$
between neighbouring backbone impurities $k$ and $j$ (there should be
no interference link between these impurities). The second product
is over interference links. We assume that each link $i$ contains
starting and final impurities $i1$ and $i3$, correspondingly,
(that are the backbone impurities) and intermediate interference
impurity $i2$. For example in our model resistor for the link connecting impurities $S$ and $3$ the index $i1$
corresponds to the starting impurity $S$. The impurities $i2$ and $i3$ are actually the scatterers $2$ and $3$ correspondingly.
The expression (\ref{chain2}) is the generalization
of (\ref{chain1}) for the arbitrary resistor.

From Eq. (\ref{chain2}) one can easily obtain the expression for
magnetoresistance corresponding to long hops and spinless
electrons in our model of independent links.
\begin{equation} \label{modMR}
\ln \frac{R(H)}{R(0)} = - \left<\sum_i \ln \frac{  \left| J_{i1}(H) +
J_{i2}(H) \right|^2}{ \left| J_{1i}(0) + J_{2i}(0)\right|^2} \right>
\approx - \overline{N}_{int}\left< \ln \frac{  \left| J_{i1}(H) +
J_{i2}(H) \right|^2}{ \left| J_{1i}(0) + J_{2i}(0)\right|^2} \right>_{int} ,
\end{equation}
where
\begin{equation} \label{J12i}
J_{i1}(H) = I_{i1,i3}(H), \quad J_{i2} =
\frac{I_{i1,2}(H)I_{i2,i3}(H)}{E_{i2}}.
\end{equation}
Here index $i$ numerates interference links; $I_{i1,i2}$, $I_{i1,i3}$ and $I_{i2,i3}$ depend on the
magnetic field through the phase $\varphi$. $\overline{N}_{int}$
on the right hand side is
 the mean number of interference links between the hopping impurities. It is proportional to the characteristic
 scatterer number $N$. Angle brackets with index $int$ mean that the averaging is over interference links
 (that correspond to interference scatterers in a resistor).

$N$ is controlled by the area which does not give  a significant addition
to the tunneling exponent. Due to the presence of backbone
impurities one should take the area with the constant width
(independent on hoping distance) $\rho \sim \sqrt{r_{sc}a} \ll
\sqrt{r_h a}$. The width $\rho$ and the distance between neighbor
impurities can be expressed as
\begin{equation}\label{rho_rsc}
\rho \sim \left( \frac{a}{n} \right)^{1/(d+1)}, \quad r_{sc} \sim \left(\frac{a^{(1-d)/2}}{ n}\right)^{2/(1+d)}.
\end{equation}
Accordingly, the mean number of scatterers $N = n r_h \rho^{d-1}$
is proportional to the hopping length $r_h$.

Finally, let us note that the quantity under the logarithm in
(\ref{modMR}) can significantly differ from unity only when
amplitudes $J_{i1}$ and $J_{i2}$ are comparable (otherwise one of
the amplitudes dominates and the discussed quantity is $\approx
1$). So we can substitute the averaging over interference links by
averaging over all scatterers and, correspondingly, substitute
$\overline{N}_{int}$ by $N$.
\begin{equation} \label{modMR1.1}
\ln \frac{R(H)}{R(0)} \approx - N \left< \ln \frac{  \left| J_{i1}(H) +
J_{i2}(H) \right|^2}{ \left| J_{1i}(0) + J_{2i}(0)\right|^2} \right> .
\end{equation}
Here averaging is over all scattering impurities $i$. In the explicit
expressions for $J_{i1}$ and $J_{i2}$ (\ref{J12i}) one should
substitute impurity $i2$ by impurity $i$. The impurities $i1$ and
$i3$ should be substituted with the backbone impurities
neighboring to $i$. With this expression the problem of
magnetoresistance in the case of long hops is deduced to the
magnetoresistance in the case of short hops for resistors with
$r_{h} \sim 2 r_{sc}$.

The expression (\ref{modMR1.1}) comes from the approximation of
the sum of all tunneling paths $\sum J_i$ and thus does not
include  free spins. To include  free spins, one should derive the
expression for magnetoresistance from (\ref{MR_Many_M}) with the
approximation of $N_{bb} \gg N_{int}$. The corresponding
expression for $\Gamma_{12}$ is derived in the Appendix
\ref{app_J_free} (eq. \ref{a1_4}). With above mentioned arguments one can
get from (\ref{a1_4}) the following expression for
magnetoresistance for long hops that includes the contribution of
free spins:
\begin{equation} \label{modMR2}
\ln \frac{R(H)}{R(0)} = - (1-P_{free}){N}\left< \ln \frac{
\left| J_{i1}(H) + J_{i2}(0) \right|^2}{ \left| J_{i1}(0) +
J_{i2}(0)\right|^2} \right> -
\end{equation}
$$
- P_{free}{N}\left< \ln \frac{ P_{\uparrow\uparrow}
\left| J_{i1}(H) + J_{i2}(H)\right|^2 +
(1-P_{\uparrow\uparrow})\left(\left|J_{i1}(H)\right|^2 + \left|
J_{i2}(H)\right|^2 \right)}{  \left| J_{i1}(0) + J_{i2}(0)\right|^2/2 +
\left(\left|J_{i1}(0)\right|^2 + \left| J_{i2}(0)\right|^2
\right)/2}\right>
$$

One can see that terms in angle brackets in (\ref{modMR2}) have
the same form as the magnetoresistance in the case of short hops
(compare with expr. (\ref{MR0free}) and (\ref{MR0fro})). This
magnetoresistance is discussed in section \ref{sect_shorthops}.
However the "length" of the "resistors" corresponding to this
terms in (\ref{modMR2}) is controlled   by the distance $r_{sc}$
between neighbor scatterers in a resistor rather than by the
actual hopping distance $r_h$. Note that $r_{sc}$ is still larger
than the distance between neighboring impurities $n^{-1/d}$.

The expression (\ref{modMR2}) means that
 all properties discussed for the short hops in section
\ref{sect_shorthops} exist also  in the case of long hops (at least as long as the independent link
approximation remains applicable). There is a negative linear magnetoresistance
due to the suppression of the interference by the magnetic field
\begin{equation} \label{MRlin}
\ln \frac{R(H)}{R(0)} \propto  - {N}(1-P_{free})
r_{sc}^{2+d/2}H \propto - r_h H.
\end{equation}
One can see that the dependence of the magnetoresistance on the
hopping distance $r_h$ (and, correspondingly, on temperature)
becomes weaker than for the short hop case (compare $r_h$ to
$r_h^{2+d/2}$). At large temperatures there is additional
temperature dependence due to a decrease of $P_{free}$ with decreasing
temperature. However, in a common situation when most
spins are frozen and $P_{free} \ll 1$, one can neglect the
dependence $P_{free}(T)$ and assume that $1-P_{free} \approx 1$.
We argue, therefore, that the interference magnetoresistance mechanism is
not affected by small concentrations of impurities with free
electron spin despite the fact that these impurities in the
limit of low temperatures are included in the hopping process.

Let us now discuss the saturation field for this negative
magnetoresistance. The conventional theory which considers the
destructive interference related to all tunneling paths leads to
the saturation field $H_{sat} \sim \Phi_0/r_h\rho$. It is the
field that suppresses coherence in a characteristic pair of tunneling
paths. Note that $H_{sat}$ tends to zero with decreasing
temperature. However, in our approximation for long hops one
considers only significant paths that in some sense are similar to
each other. Moreover, the interference phenomena are localized in
the relatively small interference links. Accordingly, the
saturation field is controlled by suppression of interference
inside one link. It leads to $H_{sat} \sim \Phi_0/r_{sc}\rho$.
This field is independent on temperature and on hopping distance.

There also exists the positive magnetoresistance due to spin
alignment by the magnetic field, which is quadratic in $H$ and has
a strong temperature dependence.  It arises from the term in
(\ref{modMR2}) related to the free spins and its expression is
\begin{equation} \label{MRord}
\ln \frac{R(H)}{R(0)} \propto   {N}P_{free} (n
r_{sc}^{(d+1)/2}a^{(d-1)/2})\left(\frac{\mu_bgH}{T}\right)^2 \propto
r_h P_{free} T^{-2} H^2.
\end{equation}
Again, this magnetoresistance mechanism has a strong temperature
dependence. If the hopping is over the Coulomb gap states, the
magnetoresistance is $\propto P_{free}T^{-2.5}$ where $P_{free}$
logarithmically decreases with temperature. The temperature
dependence of this mechanism is stronger than the one for the
other known positive magnetoresistance mechanisms including the
wave-shrinkage magnetoresistance and the magnetoresistance due to the
suppression of hops between the impurities with different
occupation numbers \cite{Kamimura,Matveev}.
 However, for the standard semiconductor with most electron
spins frozen the value $P_{free}$ is small, so this
magnetoresistance mechanism becomes important only at low
temperatures. Also one should note that this mechanism saturates
at magnetic field of the order of $T/\mu_b g$ that is also small
at low temperatures.

\subsection{numeric computation}

\begin{figure}[htbp]
    \centering
        \includegraphics[width=0.7\textwidth]{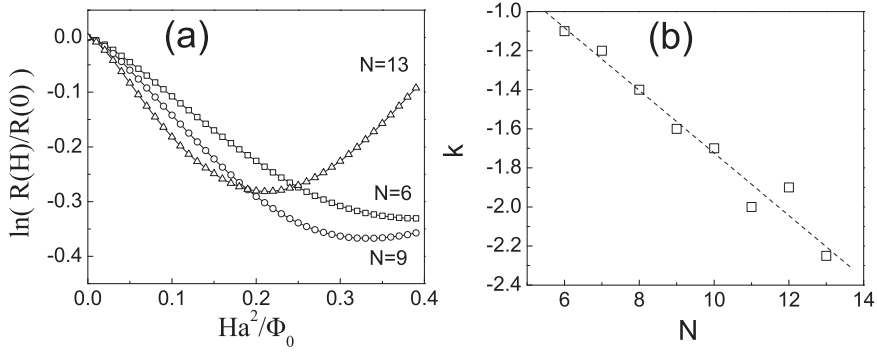}
        \caption{ (a) --- interference contribution to magnetoresistance computed with eq. (\ref{MR_Many_M}) for different
        numbers of scatterers $N\propto r_h$; (b) ---  dependance of linear expansion coefficient $k$ on the scatterer number $N$. The
        probability of an electron spin to be free was considered to be $P_{free} = 0.75$.
         }
    \label{fig:MR-R}
\end{figure}

In our analytical theory we  have used the approximation of large
variance of scatterer energy. More exactly, only when the variance
$D_E$ of $\ln I_0/|E|$ (where $E$ is impurity energy measured from
Fermi level) is large, the  model of interference links becomes
justified. Only in this case one can discriminate between
backbone, irrelevant and interference impurities and consider the
number of interference impurities to be small.

\begin{figure}[htbp]
    \centering
        \includegraphics[width=0.7\textwidth]{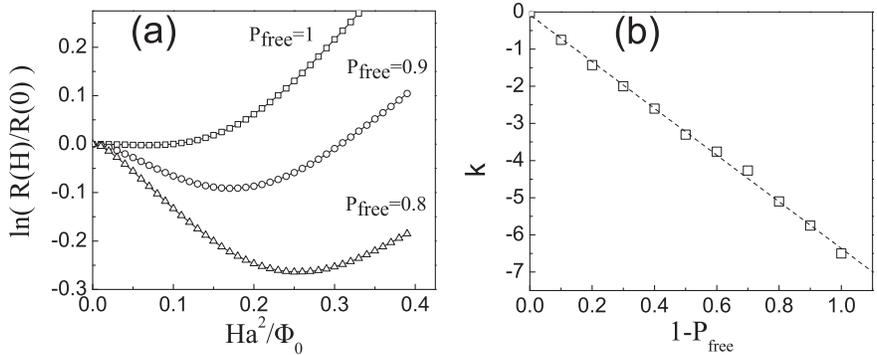}
        \caption{ Interference contribution to magnetoresistance computed with eq. (\ref{MR_Many_M}) for
        different probabilities
        $P_{free}$ (a) and  dependance of linear expansion coefficient $k$ on $1-P_{free}$
        (b). The calculation was made for resistors with $N=10$.
         }
    \label{fig:MR-P}
\end{figure}

In real semiconductors the discussed variance is of the order of
unity. However, some of our results are expected to be valid for
real semiconductors. For example, the log-normal distribution of
tunneling path amplitudes remains correct for any $D_E \ne 0$. Also we have
shown numerically that some other results (the existence of
significant pathes and backbone impurities) can be applied for realistic
distribution of scatterer energies. Consequently we should also test our
final results (\ref{MRlin}) and (\ref{MRord}) with the numeric
computations.

In our computation we simulate the random distribution of
scatterers in the area between the hopping impurities. We consider impurities only in the thin area
around the hopping line with the width $\rho \propto
\sqrt{r_{sc}a}$. The energies of the scatterers are randomly
distributed with constant density within the interval
$-I_0<E_j<I_0$. Note that this distribution corresponds to rather
small $P_{bb}$. However we show that even for this distribution
numerical results agree with our model.

 We consider the $2D$ distribution of the impurities. In our computation
we keep the impurity concentration constant ($na^2 = 0.2$), so that the
number of scatterers is directly proportional to the hopping
distance $r_h$. We assumed the hopping over the Coulomb gap states,
so that $\mu_b g/T \propto r_h^2$.

To consider the system with free electron spins, we calculate all
 the tunneling path amplitudes $J_i$ and estimate the
 magnetoresistance with a help of Eq. (\ref{MR_Many_M}). We performed the
 averaging over 1000 realizations (i.e. 1000  different critical hopping
 resistors). The results of these simulations are presented in figs.
 \ref{fig:MR-R} and \ref{fig:MR-P}. We do not use any of our approximations (essential paths, backbone scatterers)
 in our computations. Instead we include all  intermediate impurities and tunneling paths into the computation process.

 One can see that numerical
 results agree with analytic expressions (\ref{MRlin}) and
 (\ref{MRord}), i.e. there is observable linear negative magnetoresistance (even for
 large $P_{free}$); the linear expansion coefficient $k$ of magnetoresistance is  proportional to $r_h \propto N$ and $1-P_{free}$.
 Also there is a positive quadratic magnetoresistance that
 increases
 with $P_{free}$ and increases with $N$ stronger than the linear
 magnetoresistance.

 \begin{figure}[htbp]
    \centering
        \includegraphics[width=0.7\textwidth]{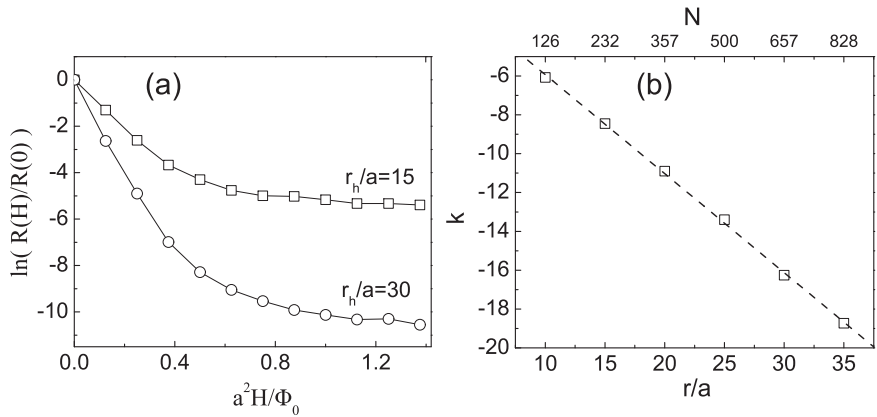}
        \caption{ The  magnetoresistance for large
        hopping distances $r_h$ and $P_{free} = 0$. (a) ---
        magnetoresistance for two different $r_h$. (b) ---
        dependance of linear magnetoresistance on $r_h$ and $N$. One
        can see that  dependance $k \propto r_h$ stays justified when the
        number of scatterers grows faster than $r_h$.
         }
    \label{fig:MR-R2}
\end{figure}

 However the question remains whether the computed linear dependence $k(r_h)$ is the real
 semiconductor property or it is due to the model choice (in our computations
 we have chosen the scatterers in the strip of constant width, and the number of scatterers was $N \propto r_h$).
 We can not consider a larger number of scatterers and compute all
 tunneling amplitudes $J_i$ because of the technical reason (the number of these amplitudes grow exponentially with
 $N$). However we can bypass this problem when we consider the system without free spins. In this case we can use the summation
 algorithm that gives the correct result (in terms of expression
 (\ref{MR_Many_M})) with the computation time $\propto N^2$.

 Accordingly we calculated the magnetoresistance for the case of $P_{free} =
 0$ and large $r_h/a \le 35$. We also took scatterers in a much
 wider area with the width $\rho \propto \sqrt{r_h}$ (so $N\propto r_h^{3/2}$). This area
 contains all scatterers that can be included in tunneling path
 without making it exponentially small (independent on backbone
 impurities). Figure
 \ref{fig:MR-R2} shows the corresponding results. One can see that the linear magnetoresistance is
 proportional to $r_h$ and not to $N$. Finally this calculation shows (fig.  \ref{fig:MR-R2} (a)) no signs of
decrease of saturation magnetic field with increasing hopping distance. This fact gives additional support to the model of interference
links.

\section{Comparison with experiment}
\label{sect_exp}

In this section we discuss one experimental result that does not
agree with the conventional picture of hopping negative
magnetoresistance. These experiments demonstrate the suppression of negative
magnetoresistance at low temperature observed in
\cite{ours5,ourPRB}. In the corresponding experiments the
magnetoresistance was measured at different temperatures in 2D
GaAs-AlGaAs heterostructures, where both the wells and the barriers
were doped by acceptor impurity Be and, thus, the acceptors within
the well included double occupied $A^+$ centers. In contrast to the
predictions of the conventional theory \cite{Book,SS,Raikh}, it
has been observed that the negative magnetoresistance is suppressed at
low temperatures.

It worth noting that the similar phenomenon has been observed earlier in
bulk semiconductors \cite{ours1,ours2}. It has been explained with the
assumption of non-Coulomb character of impurity potential at large
distances and effect of the Coulomb gap. However, as explicitly
shown in \cite{ourPRB},  this explanation does is not valid for 2D
systems (due to different asymptotic of the wavefunctions of
localized electrons in 2D with respect to 3D one). So the spin
alignment magnetoresistance mechanism has been invoked to explain this
phenomenon in 2D.  However no detailed theory of this
magnetoresistance mechanism has been proposed.

 \begin{figure}[htbp]
    \centering
        \includegraphics[width=0.6\textwidth]{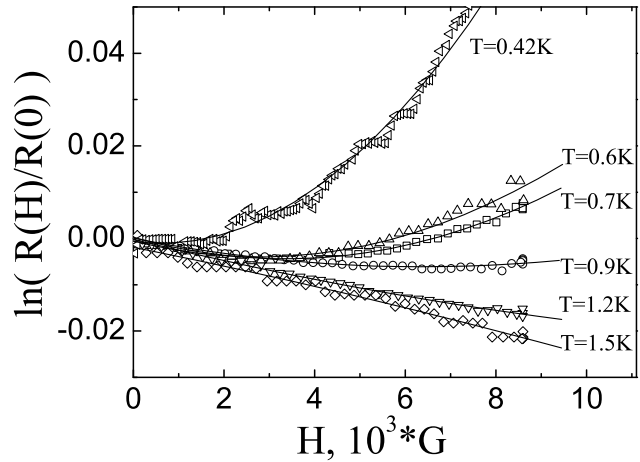}
        \caption{ Magnetoresistance of 2D semiconductor structures
        at different temperatures.
         The experiment results (dots) from \cite{ourPRB} are compared with parabolic law (\ref{parabola}) (lines).
         }
    \label{fig:exp1}
\end{figure}

It is important to note that the systems observed in
\cite{ours5,ourPRB}   include scatterers of two types (due to
the double occupation) --- the first type is the acceptor in the
quantum well that has high activation energy and a short
localization radius. The second type corresponds to the double
occupied acceptors in the well \cite{Larsen,A+} and the localized
states of the hole in the well bound to the barrier acceptors.
These states have relatively small activation energy and a large
localization radius. It is likely that the spins of holes on single
occupied acceptors in the well are free, while the spins of holes on
the scatterers of the second type are frozen. Thus, in the
considered system one has noticeable values of both $P_{free}$ and
$1-P_{free}$, and therefore both negative linear magnetoresistance and
positive spin ordering magnetoresistance should be present in the
system.

 \begin{figure}[htbp]
    \centering
        \includegraphics[width=0.7\textwidth]{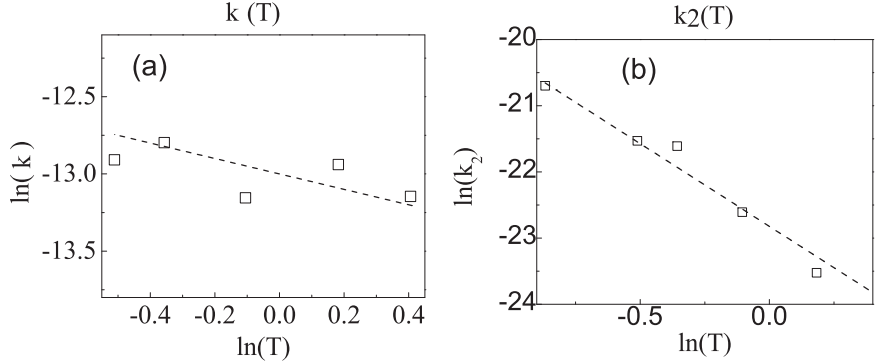}
        \caption{ The temperature dependence of $k$ (a) and $k_2$ (b) (dots) from \cite{ourPRB} compared
        to the theoretical results (\ref{MRlin}) and (\ref{MRord}) (lines).
         }
    \label{fig:exp23}
\end{figure}

Figure \ref{fig:exp1}  shows these experimental data compared
with the parabolic law
\begin{equation} \label{parabola}
\frac{\Delta R}{R} \approx \ln \frac{R(H)}{R(0)} = -kH+k_2H^2,
\end{equation}
which is  simply the first two terms of the magnetoresistance
expansion. One can see that the experimental data are in a good
agreement with (\ref{parabola}). It means that the higher terms of
expansion can be neglected when discussing these experiments.

The next step is to compare temperature dependencies of $k$ and
$k_2$ with theoretical predictions. Assuming that the conduction
is controlled by the  Coulomb gap states the conventional theory
predicts $k \propto T^{-3/2}$ and $k_2 \propto T^{-3/2}$ due to
wavefunction shrinkage magnetoresistance mechanism. These
dependencies lead to the increase of the role of negative
magnetoresistance at low temperatures. The extremal (minimal)
magnetoresistance value increases with decreasing temperature as
$\propto T^{-3/2}$. Our theory (\ref{MRlin},\ref{MRord}) for long
hops  gives $k\propto T^{-1/2}$, also we predict the strong
dependence $k_2 \propto T^{-5/2}$ due to spin alignment
magnetoresistance. The value of the magnetoresistance minimum
depends on temperature as $T^{3/2}$.

Figure \ref{fig:exp23} compares experimental data with our
results (\ref{MRlin},\ref{MRord}) demonstrating a good
agreement with our predictions. The comparison with the conventional
theory results in a significantly worse fit. So we argue
that our theory is at least in semiquantitative agreement with
experiment.

\section{discussion}
\label{sect_discus}

As we have mentioned earlier,  the conventional considerations of
the interference mechanism of magnetoresistance in the long hop
limit  \cite{PRL_perc, Medina,PRL_an,an,Spin-Orb} neglect one
important factor. It is the variance of logarithm of scatterer
energy $D_E$. In the preceding studies this parameter has been set to be
$D_E = 0$ whether explicitly or implicitly. We started our theory
from the opposite limit $D_E \gg 1$. Actually, when
$D_E = \infty$, there is no interference magnetoresistance as the
single tunneling path dominates all the hopping process. Our
theory can be considered as
 a first correction related to a finite $D_E$. One can calculate further
terms by considering sequences of two interference impurities
separated by a backbone ones, then of three interference
impurities and so on.

We did not intend to go this far in the present study. However, we presume
that the proportionality of magnetoresistance to $r_h$ should
exist also for the higher terms. Actually, for any $D_E \ne 0$
there is some probability to find a backbone impurity --- the
inclusion of this impurity significantly increase the tunneling
amplitude. So the resistor in the long hop limit is still
separated into the parts with this backbone impurities although
these parts can be large for small (but finite) $D_E$. The
magnetoresistance should be proportional to the number of these
parts and thus should be proportional to $r_h$. The dependence on
$P_{free}$ can be different in the more rigorous theory as the
parts separated by the backbone impurities can contain more than
one free spin.

One question is widely discussed in the physics of
hopping with interference in the long hop limit.  When the hops
are short, the net tunneling amplitude is usually positive. It is
not clear, however, whether the sign of net tunneling amplitude is
completely random in the long hop limit or whether a positive sign still
prevails. Our theory gives the following answer for this question.
The sign is completely random when the resistor contains at least
one backbone impurity. Naturally the net tunneling amplitude is
controlled by significant tunneling paths that include this
backbone impurity. The sign of the energy of this impurity  is
random, so the sign of the net tunneling amplitude is also
completely random. Clearly, at any temperature (and hopping
distance $r_h$) there is some chance to find a resistor without
backbone impurities but the probability of such an event drops
exponentially with $r_h$. Therefore the average sign of net tunneling
amplitude also drops with $r_h$ exponentially.

Summing up, we
generalized the theory of interference effects in the hopping
magnetoresistance to include the contribution of scatterers with
free electron spins. We considered both the case of short hops
when the mean number of scatterers in a resistor is less then
unity (many resistors in this case have no scatterers and thus no
interference effects) and the case of long hops when the mean
number of scatterers in a resistor is large. For the case of long
hops we developed a new approach to the problem of interference
magnetoresistance that is based on assumption of large variance
of the logarithm of scatter energy $D_E$ (the conventional
approaches consider $D_E=0$). We showed that our approach is
in a good agreement with numerical computations with realistic value
of $D_E$. Our theory allows one to calculate explicitly the
temperature dependence of magnetoresistance and its dependence on
the relative ratio of free spins $P_{free}$ for the short hops and
the long hops cases. Our results are in semiquantitative agreement
with experimental data on magnetoresistance in $GaAs-AlGaAs$ 2D
structures.

We are grateful to A.S. Ioselevich and N.S. Averkiev for many fruitful discussions. Our work
was supported by RFBR foundation (project n. 10-02-00544). A.V.S. also acknowledges support from "dynasty" foundation
and from the program of RAS presidium "the support for young scientists".

\appendix

\section{The expression for $\Gamma_{12}$ with the assumption of large number of backbone scatterers and with an account of free spins.}
\label{app_J_free}

To consider the problem of magnetoresistance in a system with free
spins, we should start from the expression (\ref{MR_Many_G}) for
the hopping rate. We also adopt the model of independent links,
i.e. we assume that $\Gamma_{12}$ is controlled by the significant
tunneling paths that differ from each other only in local
interference links.

First, we note that inclusion or exclusion of a scatterer with
frozen spin does not changes the tunneling result independently of
spin configuration. Each significant tunneling path $j$ has a pair
that differs from $j$ only by the interference link $i$ (for any
link $i$). One can see that if link $i$ does not contain a free
spin, then by combining such pairs one can take out the factor
\begin{equation}
 \left|I_{i1,i3} + \frac{I_{i1,i2}I_{i2,i3}}{E_{i2}} \right|^2
\end{equation}
corresponding to this interference link $i$.

Also one can take out all factors corresponding to backbone
scatterers. However, these factors do not contribute to
interference magnetoresistance, so we omit them
\begin{equation}\label{a1_3}
\Gamma_{12} \propto  \prod_{i \in frozen} \left| I_{i1,i3} + \frac{I_{i1,i2}I_{i2,i3}}{E_{i2}} \right|^2 \cdot \sum_c
P(c)\sum_{res}\left|\sum_{j(c,res)} \prod_{s \in free}
L_{s}^{(j)} \right|^2.
\end{equation}
Here the first product is taken over interference links with
frozen spin or interference links with unoccupied impurities. The
product over $s$ is taken over interference scatterers  with free
spins. $L_s^{(j)}$ is the amplitude of the interference link $s$
in the path $j$. It can be equal to either $I_{s1,s3}$ if the path
$j$ does not include scatterer $s$ or to
$I_{s1,s2}I_{s2,s3}/E_{s2}$ if the path $j$ includes $s$.

Further, we use the approximation $N_{bb} \gg N_{int}$ once
again. We consider that for any interference link $s$ with free
spin the previous impurity with free spin $s-$ is a backbone one.
Accordingly the next impurity with free spin $s+$ (for any
interference link with free spin $s$) is also a backbone impurity.

With this assumption one can see that two paths that differ only
by a single impurity with free spin $s$ lead to the same
tunneling result for all configurations when spin projections on
$s$ and $s-$ are the same. In other spin configurations this
tunneling paths lead to different results, however these results
differ only by the replacement of final projections of spin on
the impurities $s$ and $s+$. Note that in this model the inclusion
or exclusion of an interference impurity $s$ does not change the interference
pattern on the other interference impurities with free spin that
follow $s$. This interference is controlled by initial spin
projections on $s+$ and on other backbone impurities with free
spins. Thus one can obtain the following expression for
$\Gamma_{12}$.
\begin{equation}\label{a1_4}
\Gamma_{12} \propto \prod_{i \in frozen} \left| I_{i1,i3} + \frac{I_{i1,i2}I_{i2,i3}}{E_{i2}} \right|^2 \times
\end{equation}
$$
\times \prod_{s \in free} \left[
P_{\uparrow\uparrow}^{s-,s} \left|
I_{s1,s3} +
\frac{I_{s1,s2}I_{s2,s3}}{E_{s2}}\right|^2 +
P_{\uparrow\downarrow}^{s-,s}  \left(\left|
I_{s1,s3} \right|^2+ \left|
\frac{I_{s1,s2}I_{s2,s3}}{E_{s2}}\right|^2\right)
\right].
$$
Here $P_{\uparrow\uparrow}^{s-,s}$ is the probability for spins on
impurities $s$ and $s-$ to have the same projection. Naturally,
this probability does not depend on $s$:
$P_{\uparrow\uparrow}^{s-,s} = P_{\uparrow\uparrow}$ with
$P_{\uparrow\uparrow}$ defined in (\ref{PH}). Similarly,
$P_{\uparrow\downarrow}^{s-,s} = 1-P_{\uparrow\uparrow}$ for any
interference link with free spin $s$.

\end{document}